# Visualizing Atomically-Layered Magnetism in CrSBr

Daniel J. Rizzo[1], Alexander S. McLeod[1,#], Caitlin Carnahan[2], Evan J. Telford[3], Avalon H. Dismukes[3], Ren A. Wiscons[3,&], Yinan Dong[1], Colin Nuckolls[3], Cory R. Dean[1], Abhay N. Pasupathy[1], Xavier Roy[3], Di Xiao[4], D.N. Basov[1,*]

[1]Department of Physics, Columbia University, New York, NY 10027, USA

[2]Department of Physics, Carnegie Mellon University, Pittsburgh, Pennsylvania 15213, USA

[3]Department of Chemistry, Columbia University, New York, NY 10027, USA

[4]Department of Material Science and Engineering, University of Washington, Seattle, WA 98195, USA

[#]Current address: School of Physics and Astronomy, University of Minnesota, Minneapolis, MN 55455, USA

[&]Current address: Department of Chemistry, Amherst College, Amherst, MA 01002, USA

[*]Correspondence to: db3056@columbia.edu

**Abstract**

Two-dimensional (2D) materials can host stable, long-range magnetic phases in the presence of underlying magnetic anisotropy. The ability to realize the full potential of 2D magnets necessitates systematic investigation of the role of individual atomic layers and nanoscale inhomogeneity (*i.e.*, strain) on the emergence and stability of both intra- and interlayer magnetic phases. Here, we report multifaceted spatial-dependent magnetism in few-layer CrSBr using magnetic force microscopy (MFM) and Monte Carlo-based magnetic simulations. We perform nanoscale visualization of the magnetic sheet susceptibility from raw MFM data and force-distance curves, revealing a characteristic onset of both intra- and interlayer magnetic correlations as a function of temperature and layer-thickness. We demonstrate that the presence of a single uncompensated layer in odd-layer terraces significantly reduces the stability of the low-temperature antiferromagnetic (AFM) phase and gives rise to multiple coexisting magnetic ground states at temperatures close to the bulk Néel temperature ($T_N$). Furthermore, the AFM phase can be reliably suppressed using modest fields (~300 Oe) from the MFM probe, behaving as a nanoscale magnetic switch. Our prototypical study of few-layer CrSBr demonstrates the critical role of layer parity on field-tunable 2D magnetism and provides vital design criteria for future nanoscale magnetic devices. Moreover, we provide a roadmap for using MFM for nano-magnetometry of 2D materials, despite the ubiquitous absence of bulk zero-field magnetism in magnetized sheets.

## Introduction

The advent of two-dimensional (2D) materials made accessible through exfoliation of layered van der Waals (vdW) crystals has generated an explosion of research into atomically thin sheets possessing a wide range of physical properties[1,2]. Until recently[3-5], the ability to synthesize and characterize 2D materials with long-range magnetic order was notably absent. Such materials hold great promise for potential applications in spintronics[6-11], topological superconductivity[12,13], vdW heterostructures[14,15], and memory storage[16,17]. Hypothetical 2D magnets defy the Mermin-Wagner theorem[18], which states that long range magnetic order cannot exist at finite temperature within the 2D isotropic Heisenberg model[19]. Hence, magnetic anisotropy is a necessary prerequisite to realizing stable 2D magnetism, and has been demonstrated to exist in several layered materials over the last few years[3,4,20-26]. Among these, the chromium trihalides $CrX_3$ (X = Cl, Br, I) have been the most extensively studied, all of which display in-plane ferromagnetic (FM) order and either FM or antiferromagnetic (AFM) interlayer coupling depending on the choice of halide and crystal thickness[27]. Despite their extensive characterization, chromium trihalides suffer from a lack of air and moisture stability and possess relatively low transition temperatures, limiting their applications. On the other hand, CrSBr has emerged[28-31] as an air-stable layered magnetic material possessing A-type AFM ordering at comparatively high temperatures ($T_N \approx 132$ K).

A variety of analytical tools have been used to characterize CrSBr, including magnetometry[32,33], magneto-transport[32,33], Raman spectroscopy[32-34], photoluminescence (PL)[33-35], calorimetry[34], and second-harmonic generation (SHG)[34]. Using these approaches, it has been suggested that intralayer FM correlations associated with spins on Cr orbitals give rise to short-range $b$-axis polarized FM domains in each layer below a critical temperature of $T_c \approx 160$ K (i.e., well above $T_N$). This phase is dubbed the intermediate FM (iFM) phase[34], and eventually gives way to the low temperature AFM phase as interlayer correlations emerge for $T < T_N$. Reasoning based on $c$-axis spin wave confinement suggests that the difference between $T_c$ and $T_N$ decreases as the crystal thickness approaches the few-layer limit, and ultimately vanishes in the case of a single FM monolayer[34]. Theoretical modeling indicates that CrSBr has a fully spin-polarized frontier band structure[28-30], giving rise to strong coupling between excitonic transitions and magnetic order[35]. Finally, the low temperature AFM phase of CrSBr is sufficiently robust to induce proximal magnetism in bilayer graphene which subsequently displays spin-polarized conductivity and a spin-dependent Seebeck effect[36]. Despite these promising initial investigations, vital characterization of the nanoscale spatial- and layer-dependence of emergent magnetic ordering in CrSBr near its transition temperatures remains largely unexplored.

There are a variety of experimental techniques that have been used to map real-space magnetic features in 2D magnets, including scanning magneto-optic Kerr microscopy[3,4,10,22,23], spin-polarized scanning tunneling microscopy (sp-STM)[37], scanning single-spin magnetometry[38], and magnetic force microscopy (MFM)[39,40]. While measurement of the magneto-optic Kerr effect (MOKE) can provide quantitative information about the magnetic properties of a sample (such as transition temperatures and magnetization), minimum resolvable feature sizes are diffraction limited. On the other hand, sp-STM can provide atomically resolved images of the local spin density of states, but does not directly measure magnetic fields and has a limited penetration depth in multilayer samples. While scanning single-spin magnetometry is a highly sensitive, quantitative probe of magnetism down to the monolayer limit, it is technically challenging and time consuming. MFM represents an optimal compromise among these approaches, providing a suitably high lateral spatial resolution ($\geq 20$ nm) to resolve a wide range of magnetic textures[41-44] while also enabling quantitative estimates of intrinsic magnetic properties of materials with judicious modeling of probe and sample properties[45-48]. While MFM is most often used to resolve magnetic fields originating from FM materials, it can also be used to resolve magnetic fields associated with induced, stray or residual fields in AFM materials[49-52]. Despite the manifest utility of MFM for investigating magnetic behavior, its application to 2D materials has so far been minimal, being applied only to relatively thick layered magnetic materials that are effectively in their bulk magnetic state.

In our study, we employed a home-built cryogenic MFM to characterize the temperature evolution of layer-dependent magnetic contrast in few-layer CrSBr ranging in thickness from two to six atomic layers (2L – 6L). MFM directly detects nano-resolved variations ("contrast") in magnetic force $F_M$ acting between a magnetic sample and the tip of a magnetized cantilevered probe by recording frequency shifts $\Delta f \propto \frac{\partial F_M}{\partial z}$ of the cantilever resonance versus probe position (with $z$ the surface normal direction; further discussion in the Support Information, SI). These measurements were conducted at temperatures ranging from well above $T_c$ to well below $T_N$ using magnetic probes nominally magnetized to either the in-plane easy axis or the out-of-plane hard axis ($b$- and $c$-axes in Fig. 1A, respectively), affording sensitivity to fields along these axes that arise from sample magnetization. The temperature dependence of our MFM results suggests a susceptibility-dominated magnetic response at all temperatures below $T_c$. A comparatively small (but non-zero) contribution from zero-field magnetism is observed in the low-temperature AFM phase that presents as fringing fields at the edge of atomic terraces that depend on the parity of the layer thickness and step height. Meanwhile, we observed two distinct magnetic transitions associated with the onset of intralayer FM correlations at higher temperatures ($T \approx T_c$) and interlayer AFM correlations at low temperatures ($T \approx T_N$). Both transitions show a clear dependence on layer thickness and the parity of the layer number. We observe additional long-wavelength spatial variations in the magnetic transition temperatures, implying microscopic inhomogeneity in interlayer coupling that we speculate arises from local strain[53] (possibly induced by exfoliation). Finally, we selectively induced facile discrete magnetic switching events in CrSBr at temperatures close to $T_N$ using stray fields from the magnetic probe tip. Our experimental findings are closely reproduced by Monte Carlo simulations based on a Heisenberg spin lattice model. Our results lay the groundwork for quantitative magnetic imaging and nano-magnetometry of 2D materials in the few-layer limit.

**Nanoscale Magnetic Imaging of CrSBr**

Single crystals of CrSBr were grown by chemical vapor transport[33,54] and structurally characterized by single crystal X-ray diffraction (SCXRD) (Figs. 1A).[33] Few-layer flakes were isolated on $SiO_2$/Si chips using standard mechanical exfoliation techniques[2]. Their layer thicknesses were determined by topographic imaging with an atomic force microscope (AFM) (Fig. 1B). For MFM experiments, a flake with the largest distribution of layer-thicknesses was chosen to provide simultaneous collection of the MFM response from several different terraces. MFM probes with a nominal remanent magnetism of 300 emu/cm$^3$ were first coerced using NdFeB magnets with poles aligned to either the out-of-plane or in-plane direction (*i.e.*, the CrSBr $c$-axis and $b$-axis, respectively). By using MFM probes that are nominally aligned to either the hard out-of-plane or easy in-plane axes of CrSBr, we gain a multifaceted view of its magnetic response. On the one hand, the hard $c$-axis polarized probe should be sensitive to stray fields associated with the low-temperature AFM phase as well as induced magnetism caused by residual fields from the MFM probe (*i.e.*, $c$-axis susceptibility). On the other hand, the easy $b$- axis polarized probe should also be sensitive to stray fields and *b*-axis susceptibility, while providing additional sensitivity to any residual zero-field magnetism in the low temperature AFM phase. When the latter arises from a sheet of magnetic dipoles aligned to the in-plane *b*-axis, simple magnetostatic reasoning predicts negligible magnetic field over an open terrace, akin to the vanishing electric field outside a capacitor. In further analogy to the capacitor, such uniform magnetization will nevertheless generate detectable *bipolar* fringing fields near layer edges with a non-zero *b*-axis component. Extending this reasoning to out-of-plane magnetization also predicts *undetectable* magnetic force ($\frac{\partial F_M}{\partial z} = 0$) in the interior of a uniformly magnetized sheet, yet detectable fringing fields with *unipolar* *c*-axis component. For all experiments, a controlled electrical bias was applied to the MFM probe relative to the sample to compensate the contact potential difference arising from a change in work function between the tip and sample[40,55]. Using an electrically biased tip negates electrostatic contributions to our nano-resolved force measurements (see Methods for full description of our MFM measurement approach).

Using this experimental setup, we performed a series of temperature dependent MFM experiments on the CrSBr flake shown in Fig. 1B, ranging from $T = 200$ K to 40 K. Based on previous results[34], this should provide a clear view of the magnetic properties of CrSBr well above the onset of magnetic correlations (*i.e.*, the paramagnetic (PM) phase), through the iFM phase, and eventually the AFM phase at low temperature. Fig. 2A shows maps of MFM contrast collected with the tip polarization along the *b*-axis at key selected temperatures associated with the PM phase ($T > T_c$), just below the onset of the iFM phase ($T < T_c$), the onset of the AFM phase ($T \approx T_N$), and well into the AFM phase ($T < T_N$). In the colormap shown, "darker" contrasts (for decreasing $\Delta f < 0$) signify regions of qualitatively attractive magnetic force, whereas "brighter" regions imply zero or repulsive magnetic force. A complete set of MFM maps collected at all temperatures is found in Figs. S1 and S2. The distributions of the MFM contrast for each map in Fig. 2A are sorted by layer number and shown in Fig. 2B. A careful examination of these four characteristic maps yields several notable observations. In the high temperature PM phase, the MFM contrast is completely uniform across the entire field of view, indicative of a non-magnetic state. For $T \approx T_c$, a highly inhomogeneous magnetic contrast emerges with diffuse boundaries between bright (small or negligible force) and dark (large attractive force) regions. We also observe an average contrast that increases with layer number. As the temperature decreases to $T \approx T_N$, the dark contrast phase begins to shrink and is largely sequestered to odd-layer regions. Meanwhile, a bright contrast phase is observed on all layer-thicknesses that completely envelops even-layered terraces. In contrast to the behavior of the magnetic contrast near $T_c$, the boundaries between bright and dark regions are quite sharp and depend on the scanning direction (Fig. S3), indicating a nontrivial influence of the magnetic tip on the stability of the attractive, dark contrast phase. Notably, bright contrast regions at $T \approx T_c$ are correlated spatially with dark contrast regions at $T \approx T_N$, indicating a variation of the magnetic transition temperatures with position that supersedes layer number (Fig. S4). Given that strain has been shown to significantly influence the stability of magnetic phases in CrSBr[53], we speculate that this inhomogeneous spatial modulation of transition temperatures arises due to local strain induced during the exfoliation process.

The MFM map at $T < T_N$ shows spatial dependence to the contrast in the AFM phase of CrSBr that is more subtle than that of the iFM phase. The most obvious features are observed at step edges, where the polarity of the MFM contrast switches on the +b side of a given terrace compared to the –b side (Fig. S5). This polarization behaves differently for odd layers compared to even layers (*i.e.*, fringing fields on the –b side of even (odd) terraces are repulsive (attractive) while fields on the +b side are attractive (repulsive)). Such features are most prominent for step edges running perpendicular to the *b*-axis. In addition, step edges that are two layers thick (*e.g.*, the step edge between 3L and 5L) have essentially no contrast compared to single-layer steps. We interpret this contrast as arising from the presence of stray fields in the top-most layer of CrSBr that is aligned either parallel (even layer) or antiparallel (odd layer) to the in-plane easy axis (*b*-axis) and is intuitively suppressed by cancellation of fields for integral "bilayers". The observed directional, layer parity, and step-height dependence of the MFM contrast on CrSBr step edges is consistent with the expectations from an interlayer-AFM ground state.

Simultaneously, we observe a clear dependence of the interior MFM contrast on layer number in the low temperature AFM state. Notably, the addition of an odd layer increments the detectable magnetic force whereas the addition of an even layer has little effect. Hence, 3L and 4L terraces appear to have similar relative contrast, as do 5L and 6L regions. As mentioned, since a uniformly magnetized plane is not expected to give rise to detectable magnetic force directly above the sheet (*i.e.*, within the interior of a layer), one would intuitively expect little to no dependence of the MFM contrast with layer number. The overall increase of the MFM signal with layer number contradicts this naïve expectation. Hence, a magnetic response arising solely from zero-field magnetization is insufficient to explain our observations in the low temperature AFM phase of CrSBr as detected by a *b*-axis magnetic probe.

To clarify these findings, we repeated our temperature dependent study using an MFM probe polarized along the *c*-axis (Figs. 2C, D). Many of the salient features of the magnetic texture observed with the *b*-axis probe are also present in the *c*-axis data. These include: a uniform MFM contrast in the PM phase, the onset of a diffuse dark contrast (iFM) phase at $T <$ $T_c$, the presence of abrupt boundaries between dark (iFM) and bright (AFM) contrast phases at $T \approx T_N$, and a bright contrast (AFM) phase at $T < T_N$ with high contrast step edges. Despite these many similarities, there are several notable differences. For instance, the $c$-axis MFM map collected at $T \approx T_N$ records several discrete levels of contrast among both odd and even layers, whereas the corresponding $b$-axis data is much more binary with the iFM phase being restricted to odd layer regions. This is especially obvious in the 2L and 4L regions for the $T \approx T_N$ maps, where the MFM contrast is far less uniform across the $c$-axis image compared to the $b$-axis image. Nevertheless, iFM and AFM regions can still be clearly distinguished in the $c$-axis $T \approx T_N$ MFM map, with the iFM phase being primarily represented in odd-layer terraces.

In the $c$-axis image at $T < T_N$, the presence of high contrast step edges is consistent with the interpretation that stray fields associated with the edge of an AFM layer possess a nontrivial out-of-plane component. These latter findings reinforce our interpretation that analogous features observed in $b$-axis images associate with fringing fields generated at the edge of single added monolayers (Fig. S5). Unlike the $b$-axis data, the overall bulk MFM contrast observed with the $c$-axis probe in the AFM phase appears to steadily increase with layer number, despite expectations that CrSBr generates no net out-of-plane fields for any layer thickness in the interlayer-AFM ground state. Moreover, an out-of-plane component to the zero-field magnetization is not expected in CrSBr at any temperature. Therefore, the rich temperature-dependent magnetic texture observed with the $c$-axis polarized probe prompts a similar conclusion to the $b$-axis data: zero-field magnetism alone fails to account for the magnetic response captured in our MFM study.

Figs. 4A and 4B (for the $b$- and $c$-axis data, respectively) present a consolidated view of the layer- and temperature-dependent MFM maps shown in Fig. 2. These figures plot the average contrast for each layer versus temperature relative to the average contrast observed on the 2L terrace. Here the 2L response supplies a reference for all of our data that controls against transient effects within and between different MFM maps (including thermal drift in the cantilever resonant frequency, see Methods). When presented in this manner, the MFM contrast evidently increases with layer number for $T < T_c$ for both $b$- and $c$-axis polarized probes. Furthermore, the temperature-dependences of the MFM contrast for all layer thicknesses greater than 2L follows a trend that resembles the $b$-axis and $c$-axis susceptibilities measured previously for bulk CrSBr using volume-averaged magnetometry[32,33]. Taken together, the totality of the MFM data suggests that a clear understanding of the magnetic behavior of few-layer CrSBr demands an account of both zero-field magnetism and dynamically induced magnetism created by fields from our magnetic probe (*i.e.*, susceptibility).

**Extracting the Layer Susceptibility of CrSBr**

We performed a series of experiments that explore the influence of stray magnetic fields from the MFM probe on the magnetic response of few-layer CrSBr in order to address the role of susceptibility in the observed MFM contrast. We use a $c$-axis polarized probe to minimize the influence of zero-field magnetism on the measured signal. This was achieved by collecting a grid of magnetic force-distance ("approach") curves over regions of the sample possessing both dark and bright contrast phases (*i.e.*, both PM and iFM phases at $T < T_c$ or both iFM and AFM phases for $T \approx T_N$) (Figs. 3A, F). Qualitatively, one might expect nano-scale regions of the sample with a high (positive) magnetic susceptibility to produce an attractive magnetic response increasing strongly with probe-sample distance, manifesting in MFM approach curves by $\Delta f(z) < 0$ decreasing rapidly with decreasing $z$. Quantitatively, we adapt a pseudo-pole description[45] of the magnetic interaction between our realistic probe geometry and the planar few-layer sample to reproduce our approach curves. This method allows quantitatively robust extraction of the local $c$-axis magnetic sheet susceptibility $(\chi_C^{2D})$ as a single fitting parameter from the associated approach curve. Here, the sheet susceptibility $\chi_C^{2D}$ is derived from an area-normalized magnetization and thus has dimensions of length (see Supplementary discussion for complete description). We find that the spatial dependence of $\chi_C^{2D}$ largely accounts for the layer- and position-resolved collocated MFM contrast presented in Figs. 3B & G.

The first set of approach curves was obtained at $T < T_c$ (Figs. 3A–E). A high-resolution MFM map was first collected for reference (Fig. 3B) encompassing 2L, 3L, 4L and 5L regions. As observed in the large-scale image at the same temperature (Fig. 2D), a diffuse continuum of MFM contrast is observed between the iFM and PM phases. Isolating individual approach curves collected over the iFM and PM phases reveals clear differences in the associated characteristic length scale (Fig. 3A). The green curve is associated with the PM phase and the blue curve is associated with the iFM phase. The latter iFM curve has a much more rapid increase in MFM contrast with decreasing tip-sample distance than the prior PM phase, suggesting it possesses a higher value of $\chi_C^{2D}$. Applying the pseudo-pole model fit to both approach curves provides quantitative verification of this expectation, with the iFM phase showing $\chi_C^{2D} = 57$ nm and the PM phase showing $\chi_C^{2D} = 17$ nm. Repeating this procedure on a 2D grid corresponding to the location in Fig. 3B yields a spatial map of $\chi_C^{2D}$ (Fig. 3C). When comparing Figs. 3B and C, the MFM map shows essentially a one-to-one correlation with the map of $\chi_C^{2D}$. We also map the total integrated change in the MFM contrast within a given approach curve (Fig. 3D), which also acts as a qualitative proxy for $\chi_C^{2D}$ and correlates with the MFM map in Fig. 3B. Finally, we use the 2D grid of approach curves to extract a series of coarse-grained maps of the MFM signal at well-defined tip-sample separations (Fig. 3E). The MFM contrast is effectively uniform at the largest separations ($h = 200$ nm). As the tip height is reduced to $h = 50$ nm, regions with a larger value of $\chi_C^{2D}$ begin to develop MFM contrast much more rapidly than those determined to have a smaller value of $\chi_C^{2D}$. Hence, the MFM approach curves conducted at $T \approx T_c$ provide a multifaceted view of the tip-sample interaction. Our data support the conclusion that the PM and iFM phases can be distinguished in MFM by virtue of their varying susceptibility, with the MFM probe acting as a nanoscale magnetometer.

In order to explore the origin of MFM contrast between iFM and AFM magnetic phases, we repeated our approach curve analysis at $T \approx T_N$ (Figs. 3F–J). A high magnification MFM image of a region possessing 2L, 3L, 4L and 5L terraces is used as a reference for approach curves (Fig. 3G) (note that there is substantial overlap between this image and the region probed in Figs. 3A–E). Three representative approach curves are shown in Fig. 3F; one is collected with the tip held over the bright contrast AFM phase (red curve) and the other two are collected on the dark contrast iFM phase (purple and blue curves). The blue and red curves in Fig. 3F show that the iFM phase (blue) possesses a shorter characteristic length scale than the AFM phase (red). The contrasting behavior of the two approach curves indicates that the iFM phase has a higher value of $\chi_C^{2D}$ than the AFM phase (consistent with the behavior of the bulk susceptibility[32,33]). Note that a subset of the approach curves collected over the dark regions in Fig. 3G show a rapid switch at some fixed height above the sample surface (Fig. 3F, purple curve). Such curves initially behave like the AFM approach curve at large tip-sample separations, then undergo a discrete jump. After the jump, an approach profile characteristic of the iFM phase is observed. The difference between pre- and post- switching behavior can be seen clearly when extracting the best-fit values of $\chi_C^{2D}$ for the purple approach curve in Fig. 3F, isolating for regions before and after the switch (Fig. S6B). The values of $\chi_C^{2D}$ for the pre- and post-jump regions of the purple curve are 8 nm and 17 nm, respectively (Fig. S6B). These values are consistent with the values of $\chi_C^{2D}$ derived for the red (5 nm) and blue (14 nm) curves, respectively (Fig. 3F). This suggests that certain regions of the CrSBr flake that are initially in the AFM state can be coerced back into the iFM state through application of small out-of-plane fields ranging from ~300–460 Oe (*i.e.*, the residual magnetic field coming from the MFM probe, see Fig. S6).

We continue our analysis of the dependence of the MFM contrast on the underlying sheet susceptibility by extracting $\chi_C^{2D}$ across the regions shown in Fig. 3G (using the "post-jump" value of $\chi_C^{2D}$ for regions that switch with height) and plot it in Fig. 3H. Once again, the map of $\chi_C^{2D}$ in Fig. 3H largely correlates with the MFM map in Fig. 3G. One exception is in the bottom right of the images, where the MFM contrast is high but the sheet susceptibility is low. This discrepancy is due to the fact that this region has "switched" from the AFM to iFM phase in Fig. 3G, but the "post-switch" region of the associated approach curve was too small to extract a reliable value of $\chi_C^{2D}$. This forces us to fit to the "pre-switch" region possessing a value of $\chi_C^{2D}$ more akin to the bright contrast AFM phase. The integrated MFM contrast (Fig. 3I) once again behaves qualitatively similar to $\chi_C^{2D}$ and the MFM map. Hence, differences in sheet susceptibility permit us to resolve coexisting iFM and AFM phases at $T \approx T_N$, further validating the use of MFM as a nano-magnetometer.

The field- and spatial-dependence of the AFM-iFM magnetic switching is revealed by the height-dependent MFM maps derived from approach curves (Fig. 3J). These show relatively uniform MFM contrast at large tip heights ($h = 200$ nm). Regions with large $\chi_C^{2D}$ (that do not switch) show MFM contrast that increases much more rapidly with a decrease in tip height than those with small $\chi_C^{2D}$. At the same time, a dark contrast phase suddenly emerges at $h = 150$ nm that steadily grows larger in lateral dimension as the tip height is reduced to $h = 50$ nm. This process reflects the growth of the iFM phase out of an AFM ground state due to the increasing residual field of the MFM probe impinging on the CrSBr surface. Moreover, the threshold field for inducing this switch has considerable spatial dependence (Figs. 3I, S6C–E) which is likely the cause of the dependence of the dark contrast phase on the MFM scanning direction (Fig. S3). The differences in the stability of these two ground states is apparently small enough at this threshold temperature to induce switching between these states with relatively small coercive fields (as small as ~300 Oe) (Fig. S6). The AFM phase appears to be particularly unstable on odd-layer terraces, which constitute the majority of the "switching" areas observed on the sample. Thus, the stability of the AFM phases is reduced in the presence of one uncompensated AFM layer.

The coexistence of multiple magnetic phases at temperatures close to bulk $T_c$ and $T_N$ can be rationalized as a spatial dependence to these transition temperatures in the few-layer limit. To understand this, we study the evolution of the magnetic phase fractions as a function of layer number. Using the MFM maps in Fig. S2, we compute the phase fraction as a function of temperature for the PM, iFM, and AFM phases for each layer thickness. The interpolated temperature at which the PM and iFM phase fractions are equivalent is treated as $T_c$, while the temperature at which the iFM and AFM fractions are equivalent is treated as $T_N$. Fig. S7 shows the plot of $T_c$ and $T_N$ as a function of layer number, revealing that $T_c$ tends to increase with layer number while $T_N$ decreases – consistent with the prediction of ref. [34]. In addition, $T_N$ oscillates with layer number, being higher for even layers compared to odd layers. Therefore, in addition to the overall suppression of $T_N$ as more monolayers are added, the AFM phase is periodically destabilized for odd-layer terraces. We also note that there is an additional spatial dependence to the onset of the iFM and AFM phases within a given layer thickness that is consistent across multiple heating and cooling cycles (Fig. S4). Hence, layer-thickness alone cannot account for the observed spatial dependence of $T_c$ and $T_N$. Therefore, we observe multiple spatial dependences to the relative stability of the iFM versus AFM phase at $T \approx T_N$ and the iFM versus PM phase at $T < T_c$. Layer-dependence to these magnetic transitions are primarily dictated by the presence (or absence) of one uncompensated layer in odd- (or even-) layered terraces, while a concomitant layer-independent spatial modulation to these transitions is also observed (possibly arising from strain).

**Theoretical Modeling**

In order to further solidify the link between the MFM signal and the magnetic susceptibility, we use a Heisenberg spin lattice model in which calculations of the magnetic susceptibility $\chi$ may be performed directly. In our model, monolayers of CrSBr are characterized by rectangular lattices of $N_R \times N_C$ spins and $N_L$ monolayers, featuring intralayer couplings in agreement with those reported in ref. [27], which are stacked and allowed to interact *via* antiferromagnetic interlayer couplings (see Methods and section 4 of Supplementary discussion for more details). Here, we compute the sheet susceptibility along the principal $x$- ($\chi_{xx}$) and $z$-axes ($\chi_{zz}$) to provide a direct comparison to MFM data collected with $b$- and $c$-axis polarized tips, respectively. For example, $\chi_{nn}$ is defined as

$$\chi_{nn} = N_L \frac{\partial \langle M_n \rangle}{\partial B_n} = \frac{N_L}{k_B T\, N} [\langle M_n^2 \rangle - \langle M_n \rangle^2]$$

where $N = N_R \times N_C \times N_L$ is the total number of spins, and $M_n$ is the total magnetization along the $n$-axis.

The $b$-axis sheet susceptibility, $\chi_{xx}$, as a function of temperature $k_B T$ is shown in Fig. 4C. This plot shares many of the salient features of the temperature-dependent $b$-axis polarized MFM contrast shown in Fig. 4A. For all layer numbers, both plots in Figs. 4A and C gradually increase as $T$ is lowered below $T_c$ and show a peak at $T_N$ after which a comparatively sharp decrease is observed. In addition, $\chi_{xx}$ exhibits an overall increase with layer number for most temperatures that is generally in agreement with the behavior observed for the $b$-axis polarized MFM signal. Theoretically, we expect $\chi_{xx} \to 0$ as $T \to 0$; however, any off-axis contribution may generate an expected finite signal as $T \to 0$ if the experimental curve in Fig. 4A is extrapolated to $T = 0$ (see Fig. S8 for mixed $x$- and $z$-axis susceptibility). Such a situation may arise if the $b$-axis polarized tip possessed small non-trivial components along one of the hard magnetic axes, and may also be partially responsible for the observed "pairing" of MFM contrasts in the low temperature AFM phase.

The sheet susceptibility measured on the out-of-plane axis, $\chi_{zz}$, is shown in Fig. 4D, and mirrors the experimental $c$-axis polarized MFM signal shown in Fig. 4B. Again, both plots generally increase with layer number for all temperatures, and reach a maximum at the crossover temperature at which interlayer AFM correlations set in. In the low-$T$ limit, the $\chi_{zz}$ exhibits well-spaced dependence on layer number akin to the $c$-axis polarized MFM signal shown in Fig. 4B. Furthermore, $\chi_{zz}$ develops a subtle grouping behavior near the crossover point at which interlayer correlations begin to dominate. Here, the susceptibility curves of 3L and 4L nearly overlap, as do the curves of 5L and 6L – a subtle feature that is also observed in the $c$-axis polarized MFM contrast just below $T_N$. We note that a slight peak is observed in Fig. 4B at the onset of the AFM phase that is not observed theoretically. Such behavior might arise if a small component of the tip magnetic moment is inadvertently aligned to the in-plane easy axis (Fig.

S8). Our Monte Carlo simulations of the temperature- and layer-dependence of $\chi_{xx}$ and $\chi_{zz}$ in CrSBr correspond well with the experimental measures of $b$- and $c$-axis polarized MFM contrast, respectively. Therefore, we interpret our MFM data as acting primarily as a reporter of the local sheet susceptibility in CrSBr.

**Outlook**

We have performed a systematic temperature-dependent study of the magnetic properties of few-layer CrSBr using MFM. By employing probe tips magnetically polarized to both hard and easy axes of CrSBr, we were able to deconstruct the roles of stray, induced, and zero-field magnetism on the MFM contrast and correlate them to underlying magnetic ground states. This enables us to resolve the influence of single atomic layers on the stability of intrinsic and transient magnetic phases, revealing a systematic suppression of the AFM ground state in odd-layers and in the presence of modest magnetic fields. Our approach further reveals highly inhomogeneous magnetic couplings that do not correlate from layer thickness and likely arise from subtle structural distortions and strain that arise during the growth and fabrication process.

Our results show that MFM is a quantitative tool for performing nano-magnetometry on layered magnetic materials down to the 2D limit. This study provides a template for extracting figures of merit for future atomically-thin magnets and demonstrates the necessity of these spatially-resolved diagnostic tools for assessing the next generation of 2D magnetic materials. Our susceptibility-sensitive analytic approach permits direct visualization of magnetic phase growth in layered antiferromagnets, despite such systems possessing small or vanishing bulk magnetic fields. Furthermore, we provide a straightforward method for probing the behavior of highly coercive magnetic ground states by exploiting the residual field from MFM tips, revealing a promising route to tailoring nanoscale magnetic switches through control of layer parity. The insights gained in this study both significantly expand our understanding of the behavior of few-layer CrSBr and provide a conceptual foundation for the quantitative use of MFM in atomically thin magnets.

**Methods**

***Material Growth***: CrSBr was synthesized using chemical vapor transport from Cr and $S_2Br_2$ precursors following established protocols[33,54].

***Magnetic Force Microscopy***: MFM experiments were carried out on a home-built UHV cryo-AFM using commercial PPP-MFMR probes fabricated by Nanosensors™. Prior to each experiment, neodymium magnets were used to magnetize MFM probes *ex situ* in either the out-of-plane direction (*c*-axis polarized) or the in-plane fast scanning axis (*b*-axis polarized). Exfoliated CrSBr flakes were prepared for UHV experiment such that the *b*-axis of each flake was aligned to the fast-scanning axis of MFM. In order to account for small shifts in the cantilever resonant frequency ($f_0$) with changes in temperature, the drive frequency was tuned to $f_0$ prior to each measurement. A DC bias was applied to the MFM tip to minimize the frequency shift $\Delta f$ at $h = 50$ nm above the CrSBr surface to negate contributions of long-range electrostatic forces to the MFM measurement. For MFM measurements, a dual-pass approach was employed in which the first pass measures the AFM topography and the second pass measures $\Delta f$ at $h = 50$ nm above the CrSBr surface (herein, $\Delta f$ in the second pass is referred to as the MFM contrast). A median subtraction is applied to the raw MFM data to remove errant contributions to $\Delta f$ arising from slow drift in $f_0$ that occurs over the long frame time of a given MFM map (~ 2 hours). The zero-point of the MFM contrast for data presented in Figs. 2, 3, S1, S2, S3, S4, S5 and S6 is set as the largest value (*i.e.*, the least negative value) within a given map. For MFM approach curve data, heights above the CrSBr surface were defined relative to the point of "hard contact" within a given approach curve (*i.e.*, where $\Delta f$ bottoms out).

***Monte-Carlo Simulations***: We constructed a Heisenberg spin lattice model of CrSBr consisting of single-layer $N_R \times N_C$ rectangular spin lattices stacked with $N_L$ monolayers and allowed to interact *via* AFM interlayer couplings. For our simulations, $N_R = N_C = 80$ and $N_L = 2 - 6$ depending on the structure of interest. The magnetic susceptibilities were extracted from computation of the average total magnetization along a given direction *via* a single-spin perturbative Metropolis Monte Carlo (MC) algorithm. The susceptibility values are averaged over an ensemble of $20 - 50$ MC simulations per data point. A detailed description of our model can be found in section 4 of the supporting information.

**Acknowledgements**

Research at Columbia University and University of Washington was supported as part of the Energy Frontier Research Center on Programmable Quantum Materials funded by the U.S. Department of Energy (DOE), Office of Science, Basic Energy Sciences (BES), under Award No. DE-SC0019443. R.A.W. was supported by an Arnold O. Beckman Fellowship in Chemical Sciences.

**Author Contributions**

D.J.R. and A.S.M. performed all MFM experiments and analyzed the data. A.S.M. developed strategy for extracting the sheet susceptibility from MFM approach curves. C.C. performed all Monte-Carlo simulations of temperature- and layer-dependent magnetism in CrSBr. A.H.D. fabricated bulk CrSBr crystals. R.A.W. and E.J.D. prepared exfoliated CrSBr crystals for MFM experiments and performed initial characterization. Y.D. maintained cryo-AFM and assisted in troubleshooting MFM experiments. A.N.P. and C.R.D. assisted in experimental interpretation. X.R. and C.N. oversaw and advised with CrSBr growth. D.X. oversaw and assisted Monte-Carlo simulations. D.N.B. advised MFM experiments and assisted with interpretation. All authors participated in scientific discussion.

## Competing Interests

The authors declare no competing financial interests.

## Data Availability

All data presented in the manuscript are available upon request.

## Supporting Information Available

Supporting Information contains auxiliary MFM data collected with both *b*- and *c*-axis polarized tips, additional computations of the off-axis sheet susceptibility, derivation of approach curve modeling, and a detailed description of Monte Carlo simulations.

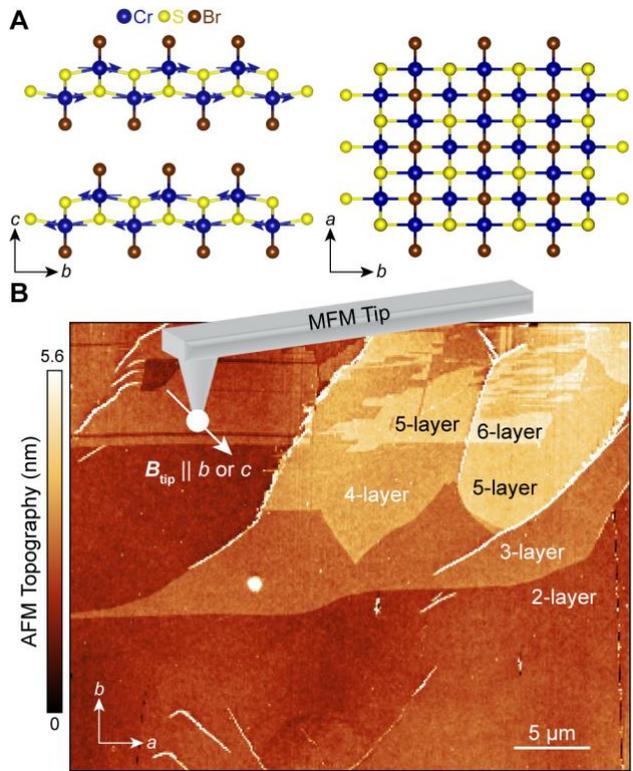

**Figure 1. MFM characterization of magnetic phases in CrSBr.** (**A**) Side- and top-down view of the crystal structure of the layered magnetic semiconductor CrSBr. Magnetic moments localized on Cr sites align ferromagnetically along the in-plane *b*-axis, and antiferromagnetically between layers along the out-of-plane *c*-axis. (**B**) AFM Topographic image of few-layer sample of CrSBr with 2 – 6-layer terraces indicated. The temperature-dependence of the MFM response of few layer CrSBr is probes with both *b*-axis and *c*-axis polarized MFM probes.

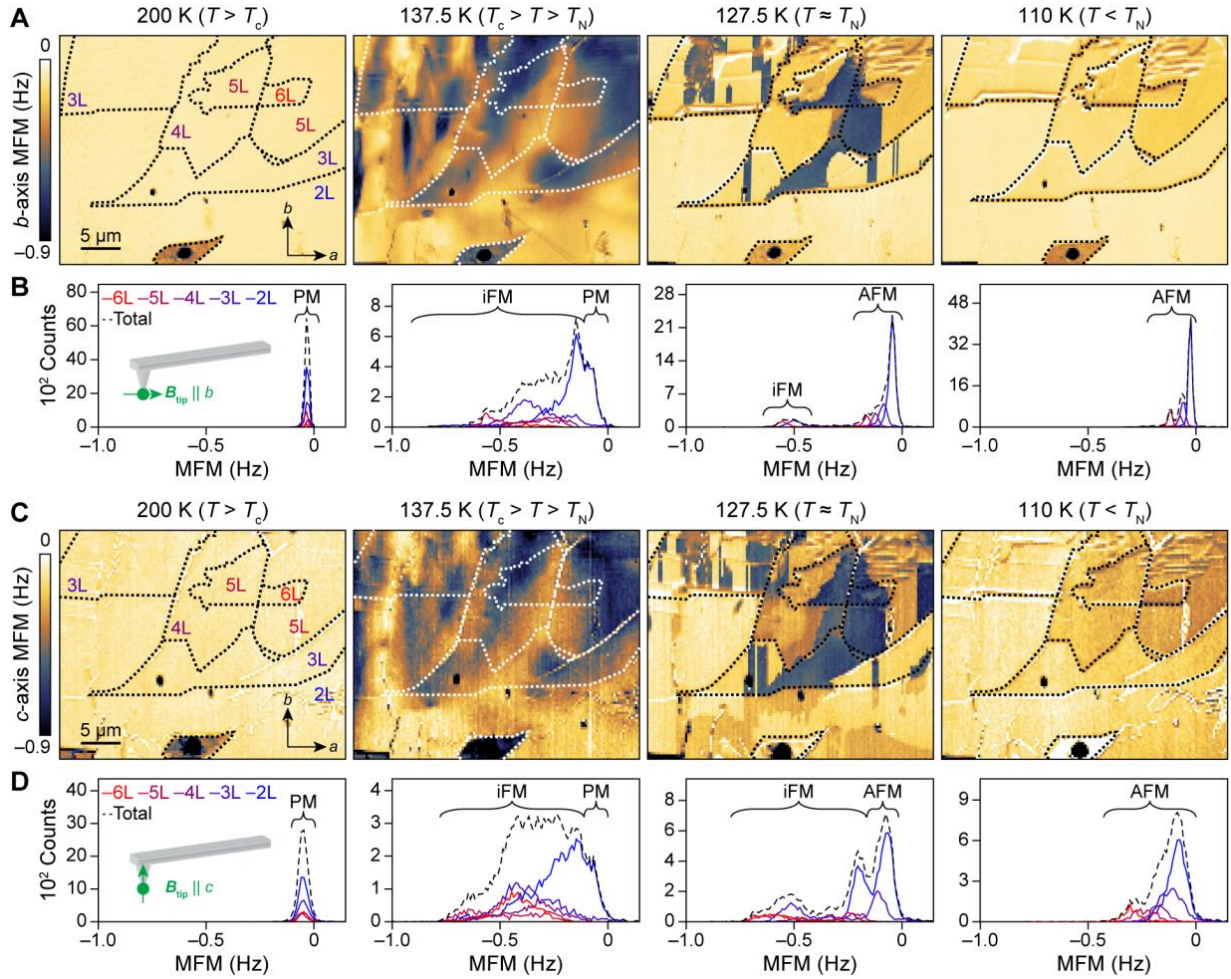

**Figure 2. Temperature-dependent evolution of magnetic phases in few-layer CrSBr observed with polarized MFM tips.** (**A**) MFM maps of few-layer (2L through 5L) CrSBr taken at four characteristic temperatures using a *b*-axis polarized tip. Uniform MFM contrast is observed for temperatures well above $T_c$ (the paramagnetic (PM) phase) which gradually give way to a diffuse, dark contrast phase at temperatures below $T_c$ (the iFM phase). As the sample temperature is brought close to $T_N$, a layer-dependent, bright contrast phase emerges characteristic of the antiferromagnetic (AFM) phase that forms sharp boundaries with the pre-existing iFM phase. The latter is primarily represented in 3- and 6- layer regions. For all temperature $T < T_N$, only the AFM phase remains. (**B**) Histograms of the MFM contrast for the four panels shown in (A). From these, it is clear that the transition from the PM to the iFM phase coincides with a continuous, broad range of MFM contrasts, while the transition from iFM to AFM creates an abrupt, discontinuous change in the MFM contrast. (**C**) The same as (A) but for a *c*-axis polarized tip. (**D**) The same as (B) but for a *c*-axis polarized tip. For $T > T_N$, the overall evolution of the MFM contrast for the *c*-axis tip is similar to that of the *b*-axis tip. However, for $T = T_N$, the *c*-axis tip additionally observes regions of intermediate contrast in the 2L and 4L regions. For the AFM phase, the dependence of the *c*-axis MFM contrast on layer number is also different than that of the *b*-axis tip.

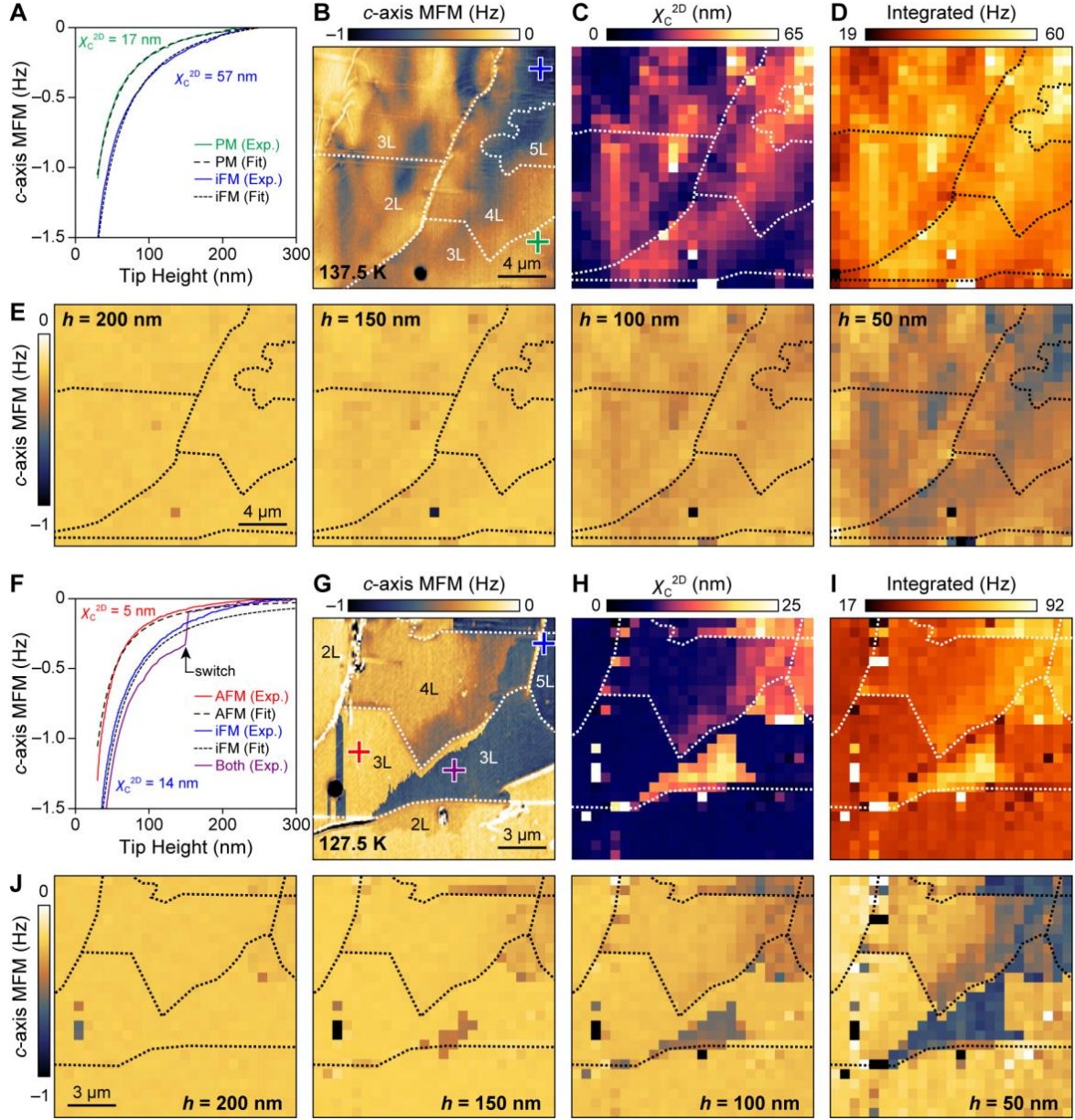

**Figure 3. Analysis of *c*-axis MFM approach curves and height-dependent MFM contrast.**
(**A**) MFM approach curves collecting over the PM (green) and iFM (blue) regions of the CrSBr near $T_c$ ($T$ = 137.5 K) as indicated by the crosshairs in (B). The PM phase has much more rapid onset in MFM contrast with tip height compared to the iFM phase. The black dashed and dotted curves correspond to the best-fit model pseudo-pole approach curve for the PM and iFM phases, respectively, with the best fit value of $\chi_C^{2D}$ indicated. (**B**) MFM image of few-layer CrSBr showing diffuse boundaries between dark and bright contrast regions (iFM and PM, respectively). (**C**) Model sheet susceptibility extracted from the approach curve map collected in the same region as (B). Regions that possess dark contrast in (B) have a larger sheet susceptibility compared to bright contrast regions. (**D**) Map of the integrated MFM contrast along each approach curve in (C). (**E**) Reconstructed height-dependent MFM maps extracted from approach curve maps, showing a gradual onset of contrast as the MFM tip-sample distance is reduced. (**F**) MFM approach curves collecting over the iFM (blue), AFM (red), and "switch" (purple) regions of the CrSBr near the Néel temperature ($T = 127.5$ K) as indicated by the crosshairs in (G). The black dashed and dotted curves correspond to the best-fit model pseudopole approach curve for the AFM and iFM phases, respectively. The purple approach curve appears to follow a similar trajectory to that of the red curve (AFM phase) for large tip-sample distances followed by a discrete jump in the MFM contrast as the tip is brought closer to the surface and subsequently resembles the blue curve (iFM phase). **(G)** MFM image of few-layer CrSBr showing sharp boundaries between dark and bright contrast regions (FM and AFM, respectively). **(H)** Model sheet susceptibility extracted from the approach curve map collected in the the same region as (G). Once again, the sheet susceptibility appears to scale with the overall MFM contrast shown in (G). **(I)** Map of the integrated MFM contrast along each approach curve in (H). **(J)** Reconstructed height-dependent MFM maps extracted from approach curve maps, showing a discrete onset, nucleation and growth of the iFM phase as the MFM tip height is reduced (and the associated tips fields impinging on the surface are increased).

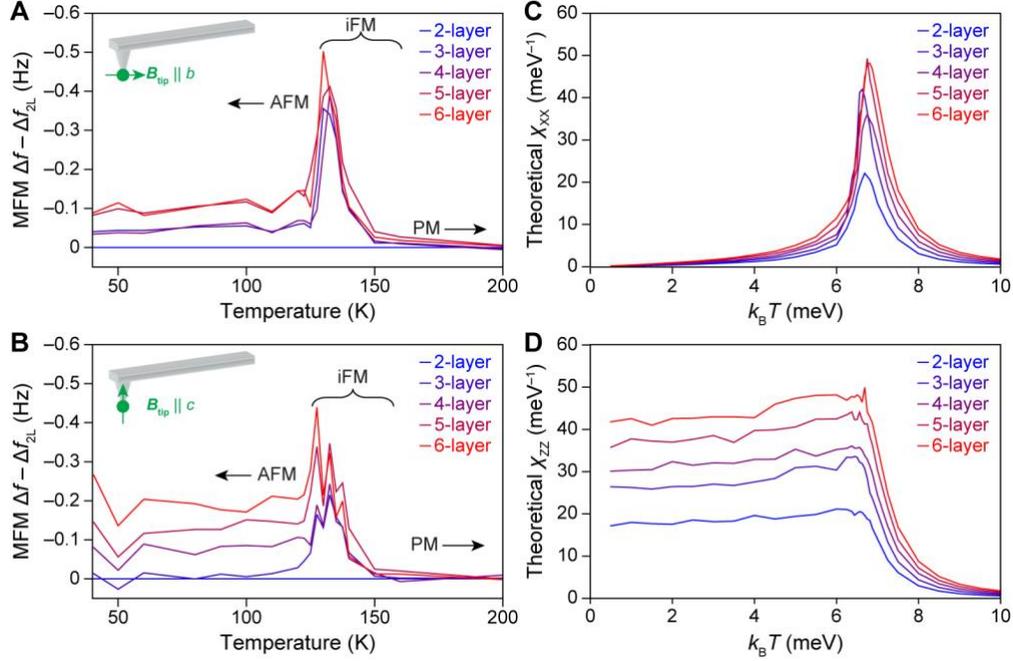

**Figure 4. Temperature-dependence of the MFM contrast compared to the theoretical sheet susceptibility.** (**A**) Temperature-dependence of the MFM contrast collected with a *b*-axis polarized MFM tip. The ground state at high temperatures is paramagnetic and yields only a small MFM contrast. As in-plane correlations begin to emerge for $T < T_c$, the MFM contrast rapidly increases and reaches a peak close to $T \approx T_N$. For $T < T_N$, the MFM contrast of the antiferromagnetic phase converges to values intermediate between the higher temperature phases and generally increase with layer number. For all temperatures, the MFM contrast generally increases with layer number. The 3L and 4L regions having similar contrast in the low-temperature AFM phase, as do the 5L and 6L regions. (**B**) Same as (A) but for an MFM tip polarized to the out-of-plane *c*-axis. The overall trend in the MFM contrast with temperature is similar to that of the *b*-axis tip, with the MFM contrast generally increasing with layer number at all temperatures. (**C**) The theoretical *b*-axis sheet susceptibility ($\chi_{XX}$) is plotted as a function of temperature. The overall temperature- and layer-dependence of $\chi_{XX}$ is similar to the MFM contrast observed in (A). (**D**) The theoretical *c*-axis sheet susceptibility ($\chi_{ZZ}$) is plotted as a function of temperature. The overall temperature- and layer-dependence of $\chi_{ZZ}$ is similar to the MFM contrast observed in (B).

**Supporting Information**

**Table of Contents:**

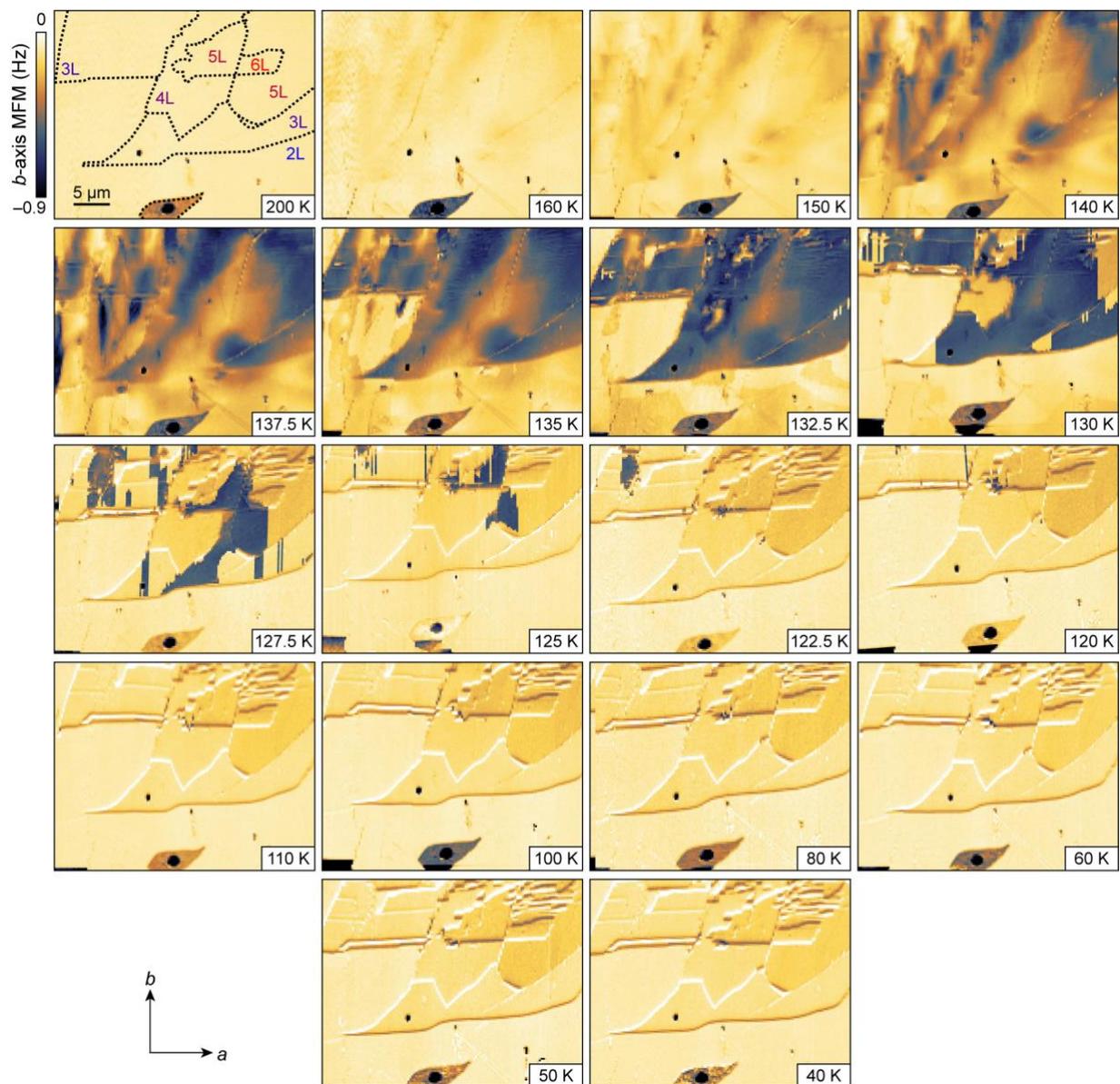

**Figure S1. Temperature-dependent MFM maps for *b*-axis polarized probe.** The MFM maps collected with the *b*-axis polarized probe labelled with the associated temperatures ranging from *T* = 200 K to *T* = 40 K.

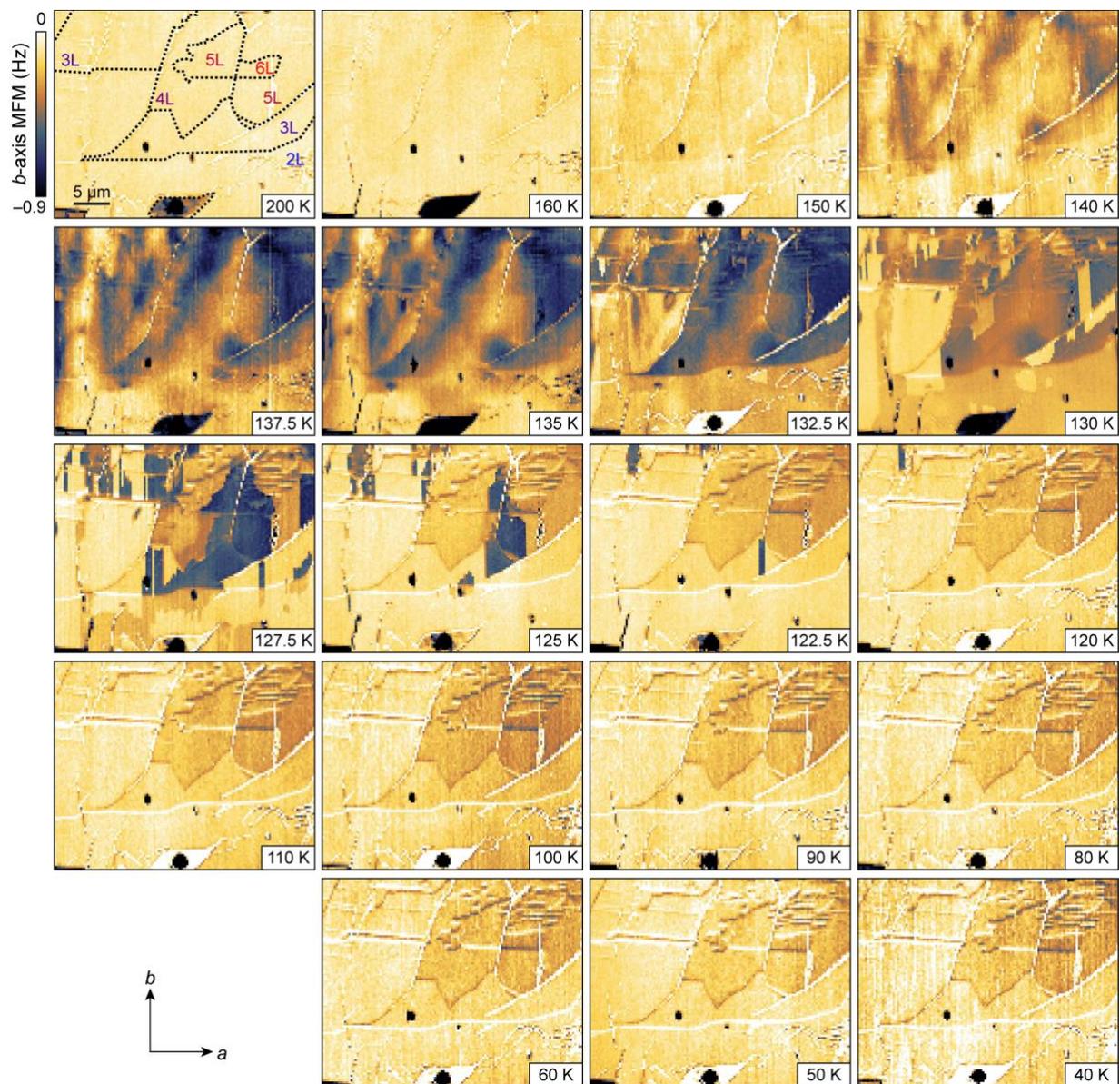

**Figure S2. Temperature-dependent MFM maps for *c*-axis polarized probe.** The MFM maps collected with the *c*-axis polarized probe labelled with the associated temperatures ranging from $T = 200$ K to $T = 40$ K.

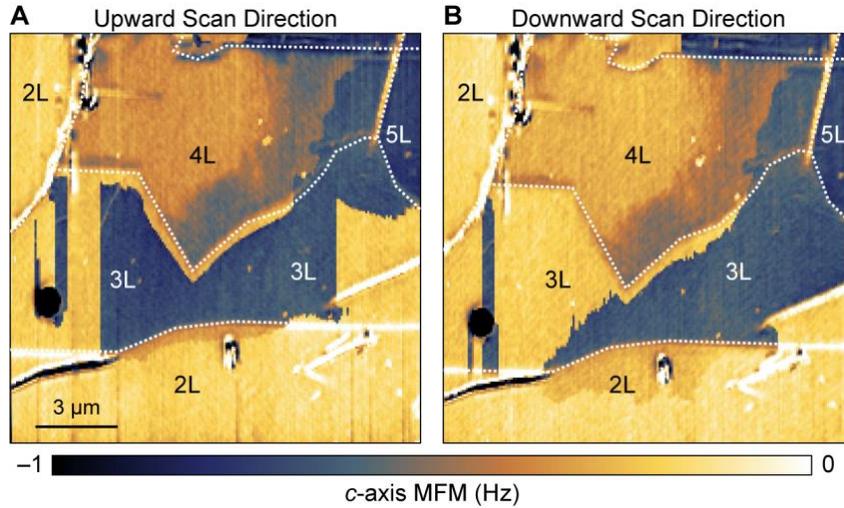

**Figure S3. Dependence of $T \approx T_N$ MFM maps on scanning direction. (A)** MFM map collected with an upward-scanning $c$-axis polarized probe at $T = 127.5$ K showing both iFM (dark blue) and AFM (yellow) phases coexisting with sharp boundaries between each phase. **(B)** The same as (A) but for a downward-scanning probe. While there is some spatial overlap between the iFM and AFM phases seen in (A), there as some regions that appear in the iFM phase in (A) that appear to be in the AFM phase in (B) (and vice versa). This demonstrates a nontrivial influence of the magnetic probe on the stability of the underlying magnetic phase.

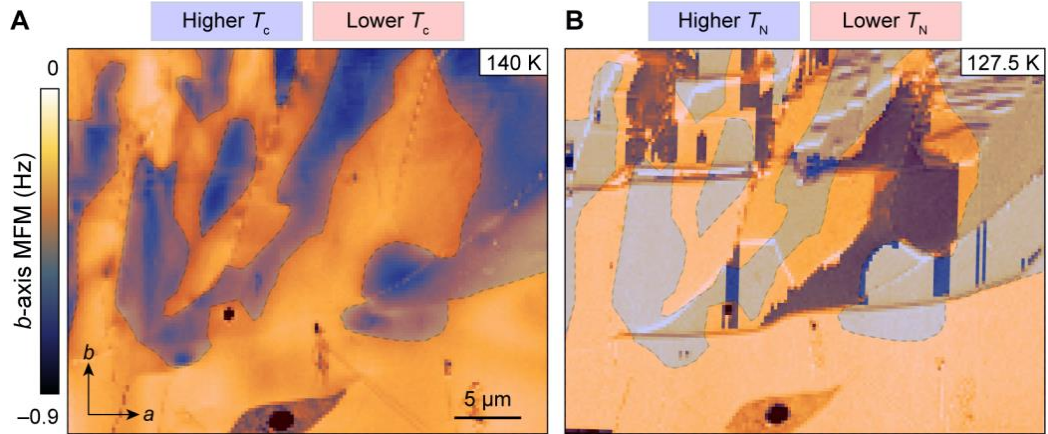

**Figure S4. Spatial-dependent phase transitions near $T_c$ and $T_N$. (A)** MFM map collected at $T$ = 140 K showing inhomogeneous spatial dependence of the yellow (PM) and dark blue (iFM) phases that cannot be understood in terms of the underlying layer number alone. Those regions that are still in the PM phase are interpreted as having a relatively low value of $T_c$ and are highlighted red, while those which are in the iFM phase have a relatively high value of $T_c$ and are highlighted blue. **(B)** MFM map collected at $T$ = 127.5 K showing inhomogeneous spatial dependence of the yellow (AFM) and dark blue (iFM) phases. Those regions that are still in the iFM phase are interpreted as having a lower value of $T_N$, while those that are already in the AFM phase have a higher value of $T_N$. The red and blue highlighted regions in (A) are reproduced in (B), showing a spatial correlation between those regions for which the PM phase persists at $T$ = 140 K and those regions where the iFM phases persists at $T$ = 127.5 K.

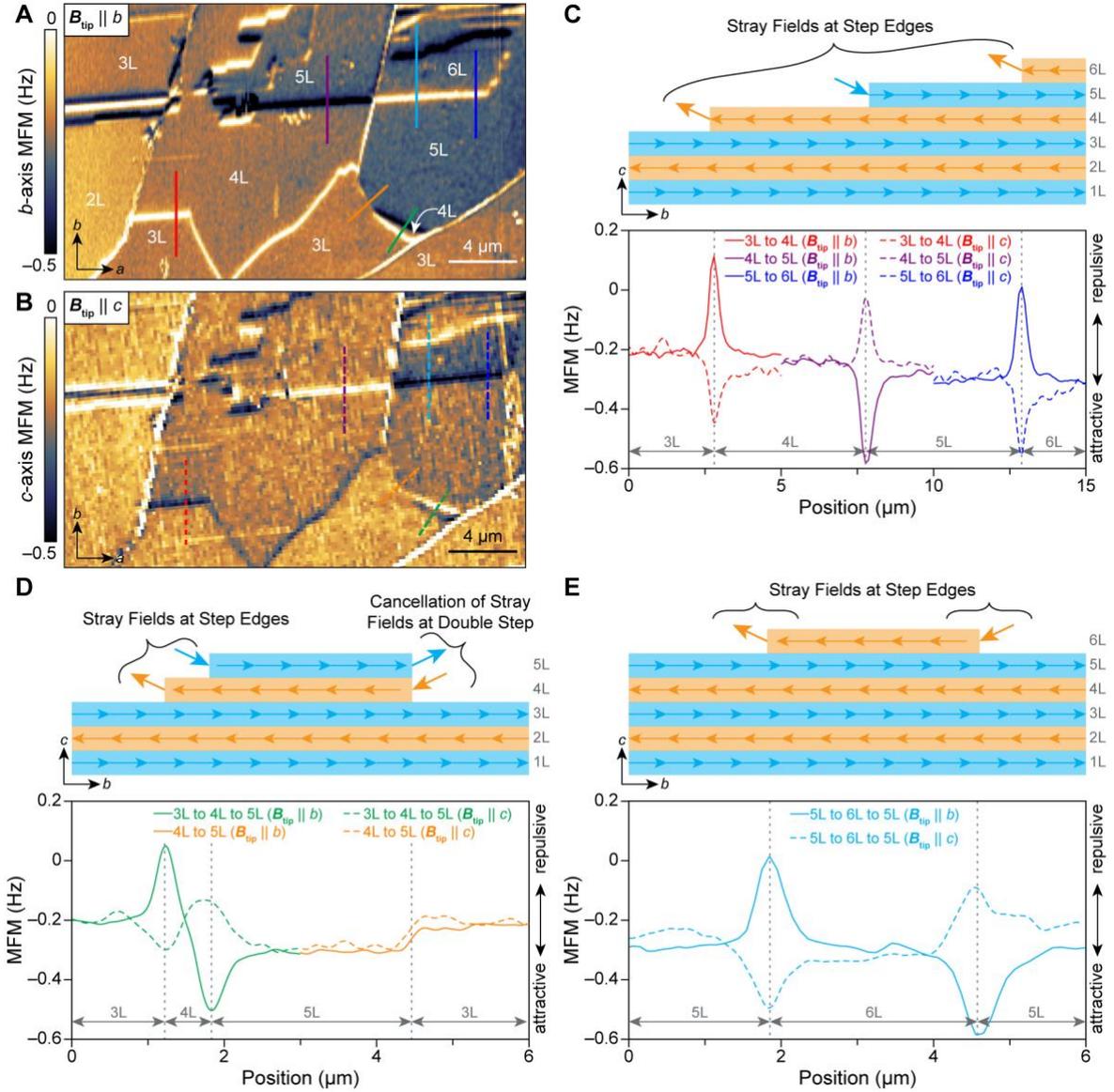

**Figure S5. Observation of stray fields on step edges in AFM ground state. (A)** MFM map collected with a *b*-axis polarized probe of the AFM phase at *T* = 33 K. The associated layer numbers are labelled. Significant bright (repulsive magnetic force) and dark (attractive magnetic force) MFM contrast can be observed at step edges. The colored lines indicate the linecuts sampled for the plots in panels (C)–(E). **(B)** Same as (A) but for a *c*-axis polarized probe of the AFM phase at *T* = 100 K. As with the *b*-axis polarized probe, significant contrast is observed at step edges. **(C)** Top panel: Schematic of fringing fields emerging at the step edges of a 3L through 6L region of CrSBr in the AFM state. Bottom panel: Plot of the spatial dependence of the MFM contrast in 3L to 4L (red curve), 4L to 5L (purple curve) and 5L to 6L (blue curve) step edges for both *b*-axis (solid curves) and *c*-axis (dashed curve) polarized probes. In all cases, a large MFM force is observed at the step edge, though the sign changes depending on the direction of the probe polarization. **(D)** Same as (C) but for a 3L to 5L region. The green curves show the spatial dependence of the MFM contrast when the 3L and 5L regions are laterally separated by a small 4L region, giving rise to large MFM contrast with opposite polarity on the two steps. The orange curves shown the MFM contrast when the 3L and 5L regions are separated

S7

by a "double step", showing no significant contrast above the bulk values due to cancellation of stray fields from the two top-most layers. **(E)** Same as (C) but for a 5L to 6L region. Here, the cyan linecut runs parallel to the $b$-axis and intersects a 5L to 6L "step up" and a 6L to 5L "step down" showing dark contrast with opposite polarity.

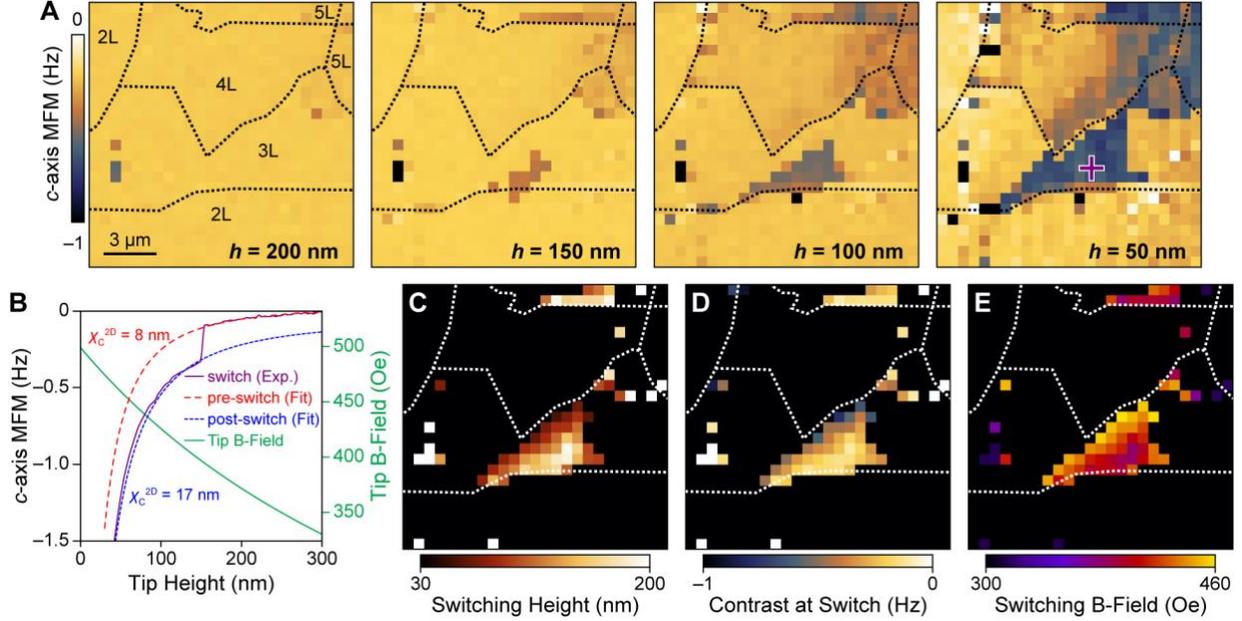

**Figure S6. Switching height map for $T \approx T_N$ approach curves. (A)** Tip-height dependent MFM maps extracted from approach curves collected with a *c*-axis polarized probe at $T = 127.5$ K (reproduced from Fig. 3J of the main manuscript). Some areas present contrast characteristic of the AFM phase at large tip-heights, but undergo a discrete switching event at some height after which the dark contrast, iFM phase is observed. As a result, the region of the maps showing the iFM phase grows in size as it is stabilized by the tip B-field at smaller tip heights. **(B)** Solid purple curve: A characteristic "switching" experimental approach curve taken in the location indicated by the cross in (A). Dashed red curve: The best-fit model approach curve to the "pre-switch" region of the purple curve, yielding a value of the sheet susceptibility characteristic of the AFM phase. Dashed blue curve: The best-fit model approach curve to the "post-switch" region of the purple curve, yielding a value of the sheet susceptibility characteristic of the iFM phase. Solid green curve: Calculated B-field generated by a pseudo-pole tip with a radius of 20 nm, tip half angle of 25°, and remanent magnetism of 300 emu/cm³. **(C)** Spatial map of the switching height for regions that are observed to undergo AFM-to-iFM transitions with decreasing tip height. **(D)** Spatial map of the MFM contrast measured at the moment the associated region undergoes an AFM-to-iFM switch. It is evident that the iFM phase is more readily induced in odd-layer numbers (3L and 5L) compared to even layer numbers (2L and 4L). However, there is also a spatial dependence to the minimum field required to induce the iFM phase that does not depend on layer number alone (*i.e.*, there is a broad range of switching heights and contrasts within the 3L region alone). **(E)** The calculated B-field shown in (B) impinging on the surface at the switching height plotted in (C).

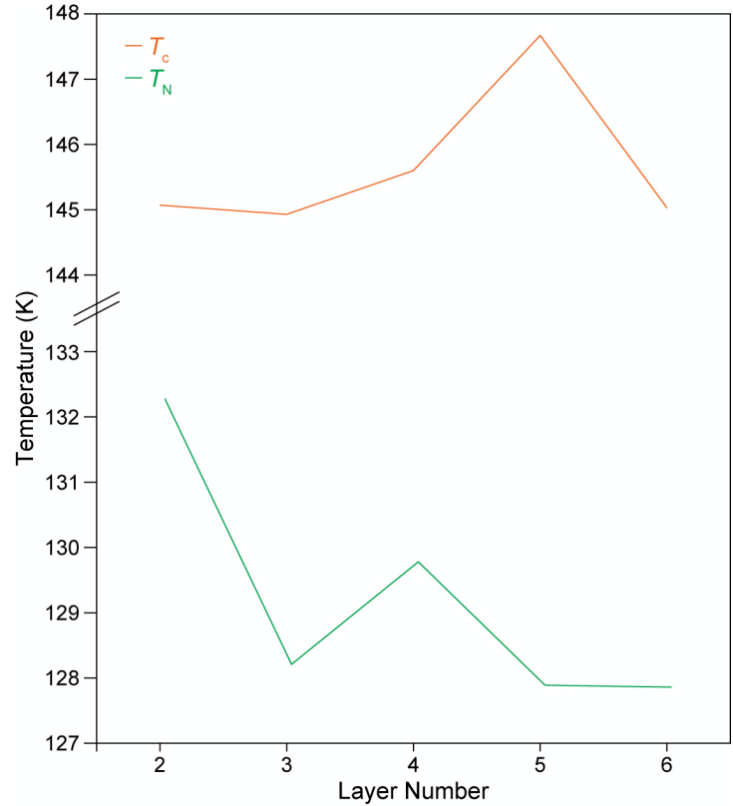

**Figure S7. Layer-dependence of $T_N$ and $T_c$.** The experimental values of $T_c$ (orange line) and $T_N$ (green line) are plotted as a function of layer number. Here, $T_c$ is defined as the interpolated temperature at which phase fraction of PM and iFM phases are equal for within a given layer, and $T_N$ is define as the interpolated temperature at which iFM and AFM phases are equivalent. $T_c$ tends to increase with layer number while $T_N$ tends to decrease with layer number. Odd-layer thicknesses experience an additional suppression of $T_N$ compared to even layers.

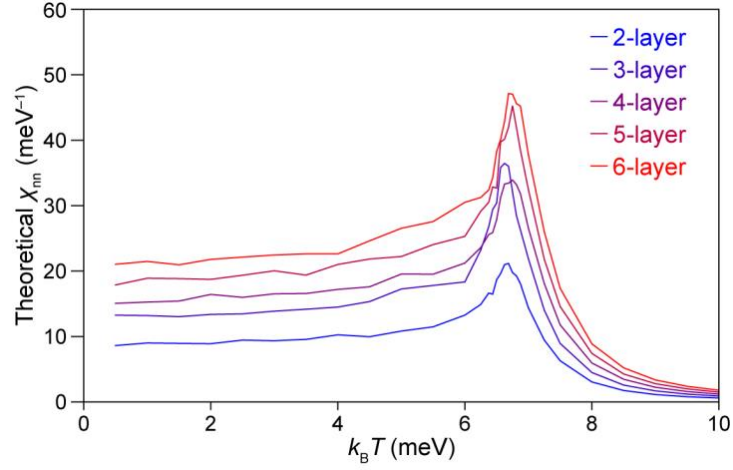

**Figure S8. Theoretical off-axis sheet susceptibility.** Theoretical temperature- and layer-dependence of the sheet susceptibility for magnetization sampled at $\hat{n} = \frac{1}{\sqrt{2}}[\hat{x} + \hat{z}]$ (*i.e.*, 45° between *b*- and *c*-axis directions). This theoretical data provides a reference for potential off-axis components to $B_{\text{tip}}$ when interpreting the data in Figs. 4A, B.

**Supplementary Discussion**

**Section 1. Magnetic transfer function of a magnetized probe**

The magnetic potential $\phi$ associated with an arbitrary bounded distribution of magnetization described by $\boldsymbol{M}(\boldsymbol{r})$ is given by:

(1)
$$\phi(\boldsymbol{r}) = \frac{1}{4\pi} \int_\Omega dV' q_m(\boldsymbol{r}') \frac{1}{|\boldsymbol{r}-\boldsymbol{r}'|} = -\frac{1}{4\pi} \int_\Omega dV' \, \boldsymbol{\nabla} \cdot \boldsymbol{M}(\boldsymbol{r}') \frac{1}{|\boldsymbol{r}-\boldsymbol{r}'|},$$

where $\Omega$ denotes the volume occupied by finite $\boldsymbol{M}$, and $q_m$ denotes the effective "magnetic charge" associated with divergence in magnetization. If the magnetization is axisymmetric about an axis $z$, then in cylindrical coordinates $(\rho, z)$ the divergence can be rewritten and the "Coulomb kernel" can be expanded in a basis of cylindrical waves:

$$\phi(\boldsymbol{r}) = -\frac{1}{4\pi} \int_\Omega d\rho' dz' \left( \frac{1}{\rho'} \frac{\partial}{\partial \rho'} (\rho' M_\rho) + \frac{\partial M_z}{\partial z'} \right) \rho' \int_0^\infty dq \, J_0(q\rho') \, J_0(q\rho) \, e^{-q|z-z'|}$$

When evaluated at positions $z \equiv -z_p < 0$ below a magnetized distribution for which only $M_z$ is nonzero for $z' > 0$, the expression simplifies:

(2)
$$\phi(\boldsymbol{r}) = -\frac{1}{4\pi} \int_0^\infty dq \, J_0(q\rho) \int_\Omega dz' \, d\rho' \rho' \, J_0(q\rho') \, e^{-q|z-z'|} \frac{\partial M_z}{\partial z'}$$
$$\equiv \frac{1}{4\pi} \int_0^\infty dq \, q \, J_0(q\rho) \, \phi_p(q) \, e^{-qz_p}$$

Here we identify $\phi_p(q)$ as the "transfer function" of the magnetization relative to $z = 0$, and the subscript $p$ indicates that forces on the magnetization functionalize it as a magnetic probe. Up to an exponential propagator in the probe-sample separation distance $z_p$, $\phi(\boldsymbol{r})$ is then given by the Hankel transform of $\phi_p(q)$, and the latter is readily identified as:

(3)
$$\phi_p(q) \equiv -\frac{1}{q} \int_\Omega dz' \, d\rho' \rho' \, J_0(q\rho') \, e^{-qz'} \frac{\partial}{\partial z'} M_z(\rho', z')$$

This expression is itself a weighted Hankel transform over variations in the magnetization, otherwise regarded as the distribution of "magnetic charge".

Eq. (2) implies that for a magnetic transfer function $\phi(q) \propto q^N$, the associated magnetic potential at $\rho = 0$ is likewise a power law $\phi(z_p) \propto z_p^{-(N+2)}$ for $N > -2$. Noting that a magnetic $M$-pole (*e.g.*, a dipole corresponds to $M = 2$) exhibits $\phi_M(z_p) \propto z_p^{-M}$, we conclude that an $M$-pole is described by a transfer function $\phi_M(q) \propto q^{M-2}$. On the other hand, a distribution $M_z(z') \propto z'$ is integrable in Eq. (3) over a finite-sized probe volume $\Omega$, *e.g.*, over an interval $0 \leq z' \leq L$, to the marginal case described by a transfer function with $N = -2$, but only in the limited interval $L^{-1} < q < \infty$. Integrating Eq. (2) over this valid range of momenta yields $\phi(z_p) \propto \log(z_p/L)$ for $z_p \ll L$. The associated magnetic field is described by $H_z \propto -\partial_{z_p}\phi \propto 1/z_p$, a so-called "pseudo-pole" field[1]. As described in the next section, this case is directly relevant to a physical model for the realistically extended magnetic force microscopy probe used in the present experiments.

## Section 2. Transfer function of a realistic probe geometry: the magnetized hyperboloid

A semi-infinite magnetic probe of conical geometry with an opening half-angle of $\theta$ can be approximated by a hyperboloid of revolution with a radius of curvature $a$ at the apex. The transfer function for such a hyperboloid with metallic coating (thickness $t$) magnetized in the z-direction (with magnetic moment per unit volume $M$) can be deduced approximately as follows. We parameterize the inner and outer surfaces of the probe's magnetic coating by functions $z_{\pm}(\rho)$, which denote lower and upper height coordinates $z$ corresponding to a single radial coordinate $\rho$, where $\rho = 0$ defines the probe (z-)axis. For simplicity, we will assume $z_+(\rho) \approx z_-(\rho) + t$. The hyperboloidal profile is described by:

(4)
$$z_-(\rho) = \sqrt{(\rho/\tau)^2 + (a/\tau^2)^2} - a/\tau^2$$

where $\tau \equiv \tan\theta$. For simplicity, we henceforth consider $z$ and $\rho$ in units of $a$. For a uniformly magnetized coating, $\boldsymbol{M} = M\,\hat{z}$ is constant between surfaces $z_{\pm}(\rho)$, and we have:

$$-\frac{\partial}{\partial z'}M_z(\rho',z') = M\left[\delta\big(z' - z_+(\rho')\big) - \delta\big(z' - z_-(\rho')\big)\right].$$

In the limit that $\delta \ll a$ we can approximate the transfer function defined in Eq. (3) by:

(5)
$$\phi(q) \approx \frac{tM}{q}\int_0^\infty d\rho'\,\rho'\,J_0(q\rho')\left[\frac{d}{dz'}e^{-qz'}\right]_{z'=z_-(\rho')}$$

$$= -tM\,e^{q/\tau^2}\int_0^\infty d\rho'\,\rho'\,J_0(q\rho')\,e^{-q\sqrt{(\rho'/\tau)^2+\tau^{-4}}}$$

$$= -tM\,\tau^2\,e^{q/\tau^2}\int_0^\infty dx\,x\,J_0(q\tau x)\,e^{-q\sqrt{x^2+\tau^{-4}}}$$

$$= -tM\,\frac{e^{q/\tau^2}}{\sigma^3}\left(\frac{\tau^4 + \sigma\,q}{\tau^2 q^2}\right)e^{-q\,\sigma/\tau^4}$$

$$= -tM\sin^2\theta\left(\frac{\cos\theta + \cot^4\theta\,q}{q^2}\right)\exp[-q\cot^2\theta\,(1+\cot\theta\csc\theta)]$$

Here we have used the change of variable $x \equiv \rho'/\tau$ in order to apply Eq. (2.12.10.7) of ref. [2], and we have defined $\sigma \equiv 1/\cos\theta$.

We can draw several qualitative observations from Eq. (5). For a narrow hyperboloid the cotangent in parentheses dominates and, below an exponential cutoff in momentum $q$, we find $\phi(q) \propto q^{-1}$ characteristic of a magnetic monopole, as one might obtain from a column of vertically stacked dipoles. The squared sine pre-factor reflects that the volume of the probe vanishes as $\theta \to 0$ and so also does its magnetic field. For an oblate hyperboloid $\theta \to \pi/2$, the cosine in parentheses dominates and carries the "pseudo-polar" contribution $\phi(q) \propto q^{-2}$ as one might obtain from a column of vertically stacked monopoles[1]. By dimensional analysis, this feature is generic to a distribution of axial dipoles whose density grows in proportion to their distance, yielding a magnetic potential that is logarithmic in the proximal probe-sample distance $z_p$, and a magnetic field scaling unconventionally as $z_p^{-1}$. The cosine vanishes in the limit $\theta = \pi/2$, which matches expectation that the external field will vanish from a homogeneous planar surface of magnetic dipoles. Whereas the radius of curvature $a$ (the probe "sharpness") is here held fixed irrespective of $\theta$, the momentum cutoff nevertheless shows a surprising $\theta$-dependence: As the oblate hyperboloid tends towards a plane at $\theta \to \pi/2$, the exponential cutoff softens at high momentum – a synthetic "sharpening" effect that counteracts the reduced transfer at high $q$ from the "pseudo-pole". A compromise is reached between increasing momentum transfer (softening momentum cutoff) and the overall falling field strength ($\propto \cos \theta$) upon increasing half-angle up to (and not beyond) $\theta \approx 1$, or about $\theta \approx 60°$. For such choice of conical angle, the oblate probe might in principle supply best-resolved spatial sensitivity to local magnetic fields. A conventional probe geometry such as that used in the present experiments is reasonably described by $\theta \approx 30°$, for which $q^{-1}$ and $q^{-2}$ contributions to the probe's magnetic potential should be considered equally relevant for any quantitative analysis.

## Section 3: Force between a magnetic probe and a magnetized material

We now consider the $z$-magnetized MFM probe as such an axisymmetric distribution of magnetization, as comprised by the magnetic coating over the probe surface. In the case where the probe's field magnetizes a planar sample (whose surface we now place at $z = 0$) with permeability $\mu$, the induced magnetic field will produce a force on the probe. We can compute this force $F_z$ between the probe and sample by considering the total magnetic energy $\mathcal{E}_m$. In the following $\boldsymbol{B} = \boldsymbol{B_p} + \boldsymbol{B_s}$ is the total magnetic field summing contributions from the probe and sample, respectively, whereas $\boldsymbol{H} = \mu\boldsymbol{B} = -\nabla(\phi_p + \phi_s)$ for probe- and sample-derived magnetic potentials $\phi_{p,s}$, and $z_p$ denotes the closest probe-sample separation distance:

(6)
$$\mathcal{E}_m = \frac{1}{4\pi} \int dV \, \boldsymbol{B} \cdot \boldsymbol{H} \quad \Rightarrow \quad F_z = -\frac{\partial}{\partial z_p} \mathcal{E}_m$$

$$\therefore \, F_z = -\frac{1}{4\pi} \frac{\partial}{\partial z_p} \int dV \, |\nabla \phi|^2 / \mu$$

$$\approx -\frac{1}{2\pi} \frac{\partial}{\partial z_p} \int dV \, \nabla \phi_p \cdot \nabla \phi_s / \mu$$

$$\approx \frac{1}{2\pi} \frac{\partial}{\partial z_p} \left( \int dV \left( \phi_s \, \mu^{-1} \nabla^2 \phi_p + \phi_s \, \nabla \mu^{-1} \cdot \nabla \phi_p \right) - \int d\boldsymbol{A} \cdot \phi_p \nabla \phi_s / \mu \right)$$

The approximation involves discarding the $z$-independent contribution to energy associated with $\boldsymbol{B_p}$ and excluding terms scaling as $|\nabla \phi_s|^2$ through the assumption that $|\boldsymbol{B_s}| \ll |\boldsymbol{B_p}|$. The last step applies integration by parts. The second term in Eq. (6) vanishes when the domain(s) of volume integration have piecewise homogeneous permeability. On the other hand, taking the volume integral over all space, the surface integral (element $d\boldsymbol{A}$ oriented "outward") vanishes at infinity, leaving $\nabla^2 \phi_p = -4\pi \nabla \cdot \boldsymbol{M}_p$ nonzero only within the probe's volume. Meanwhile, taking the region(s) where $\mu \neq 1$ to be "small" compared to variations in $\phi_{p,s}$, as when the sample is a thin magnetic layer, the second term in the volume integral likewise tends toward vanishing. Then, applying integration by parts twice in sequence over the half-space volume $z \geq 0$ (including the probe) where $\mu = 1$, we obtain:

(7)
$$\begin{aligned} F_z &\approx -\frac{1}{2\pi} \frac{\partial}{\partial z_p} \left( \int_\Omega dV \, \nabla \phi_s \cdot \nabla \phi_p + \int_{z=0} dA \, \hat{z} \cdot \phi_s \nabla \phi_p \right) \\ &\approx \frac{1}{2\pi} \frac{\partial}{\partial z_p} \int_{z=0} dA \, \hat{z} \cdot \left( \phi_p \nabla \phi_s - \phi_s \nabla \phi_p \right) \\ &\approx \frac{1}{2\pi} \frac{\partial}{\partial z_p} \int_{z=0} dA \left( \phi_p \frac{\partial}{\partial z} \phi_s - \phi_s \frac{\partial}{\partial z} \phi_p \right) \end{aligned}$$

Here we have applied the defining feature of the response field: $\nabla \cdot \boldsymbol{H}_s = -\nabla^2 \phi_s = 0$. Thus, Eq. (7) evaluates the probe-sample force strictly through a surface integral at $z = 0$ in terms of the "driving" field $\phi_p$ and the response field $\phi_s$ "reflected" from the sample, and their surface-normal derivatives.

The assumed axisymmetry allows an angular spectrum representation of fields in the probe-sample gap, including the vicinity of $z = 0$, via their Hankel transform $\phi_{p,s}(q)$:

(8)
$$\phi_{p,s}(\boldsymbol{r}) = \int_0^\infty dq \, q \, J_0(q\rho) \, e^{-q(-z_p \mp z)} \phi_{p,s}(q)$$

Whereas $\phi_p(q)$ is simply the probe transfer function, the assumed planar sample geometry also implies that $\phi_s(q) = -r_\mu(q)\phi_p(q)$, for some momentum- and permeability-dependent "magnetic reflection coefficient" $r_\mu$. Combining Eqs. (7) and (8) and utilizing the areal integral relation below yields:

(9)
$$\int_{z=0} dA \, J_0(q'\rho) \, J_0(q\rho) = \frac{2\pi}{q} \delta(q - q')$$
$$\therefore \ F_z \approx -2 \int_0^\infty dq \, q^3 \, r_\mu(q) \, \phi_p(q)^2 \, e^{-2qz_p}$$

We can observe some basic phenomenology from Eq. (9) by assuming for demonstration a constant (momentum-independent) magnetic reflectance. The transfer function of an $M$-pole probe $\phi_p \propto q^{-(M+2)}$ admits analytic evaluation of Eq. (9), yielding a force $F_z \propto \left(2z_p\right)^{-2M}$, as one might expect from two $M$-poles interacting over a distance $2z_p$. For the "pseudo-polar" case $M = 0$, the force is again logarithmic in the normalized distance $2z_p/L$, with $L$ the characteristic size of the probe.

Note that for positive "reflectance" $r_\mu$, the probe-sample force $F_z < 0$ is attractive. The correspondence $\mathbf{D} = -\varepsilon\nabla\phi \Leftrightarrow \mathbf{H} = -\mu\nabla\phi$ between electrostatics and magnetostatics allows to repurpose formulas for the electrostatic reflectance among layered media to obtain the corresponding magnetostatic reflectance simply by interchanging $\varepsilon$ with $\mu$. To wit, the magnetostatic reflectance of a single layer of non-unity permeability $\mu$ and finite thickness $d$ is given by:

(10)
$$r_\mu(q) = \frac{r_1 + r_2\,e^{-2qd}}{1 + r_1 r_2\,e^{-2qd}} \quad \text{with} \quad r_1 = \frac{\mu-1}{\mu+1} \quad \text{and} \quad r_2 = -r_1$$

This form is easily obtained by the transfer matrix method, as applied commonly to electrostatics of layered media. Eq. (10) has the general characteristic of "transmitting" fields of momenta $q \ll d^{-1}$ as $r_\mu \propto q$ and tends asymptotically to $r_1$ for $q \gg d^{-1}$. Therefore an $M$-pole probe will transition between two distinct power laws as $z_p$ approaches $d$, whereas $\mu$ sensitively controls the onset of this transition and thereby the overall shape of the force-distance curve, as well as the overall magnitude of the magnetic force. A magnetic probe interacting with a CrSBr layer only a few nanometers or less in thickness corresponds to the limit that $q \ll d^{-1}$, in which case Eq. (10) simplifies to:

(11)
$$r_\mu(q) \approx \frac{q}{q - q_\chi} \quad \text{with} \quad q_\chi \equiv -(2\pi\chi d)^{-1} \quad \text{where} \quad \mu = 1 + 4\pi\chi$$

It is clear that the factor $\chi d$ indeed represents an emergent length scale and is referred to as the sheet susceptibility ($\chi_c^{2D}$) in the main manuscript. A thin magnetic film will polarize in response to fields confined below this length scale, while transmitting fields that are comparatively delocalized. Thus, for a prescribed magnetic probe geometry and momentum transfer function, the dimensionless product $q_\chi a$ determines the shape of the force-distance curve with respect to the dimensionless distance $z_p/a$, with a secondary influence on the overall amplitude of the force, whereas factors like the probe magnetization exclusively scale its amplitude. We consider Eq. (11) to faithfully describe the magnetic response of our few-layer CrSBr flakes.

The measurable quantity by our magnetic force microscopy experiments is the frequency shift $\Delta f = f(z_p) - f_0$ of the probe's cantilever resonance relative to its unloaded frequency $f_0$. For probe tapping amplitudes small relative to the length scale of magnetic features, this shift is proportional to the $z_p$-derivative of the magnetic force (Eq. (7))[3]:

(12)
$$\Delta f(z_p) \approx -\frac{f_0}{2k}\frac{\partial}{\partial z_p}\left[-2\int_0^\infty dq\; q^3\, r_\mu(q)\,\phi_p(q)^2\,e^{-2qz_p}\right]$$
$$\approx -\frac{2f_0}{k}\int_0^\infty dq\; q^4\, r_\mu(q)\,\phi_p(q)^2\,e^{-2qz_p}$$

Here $k$ is the spring constant of the probe's cantilever. Note that while the probe-sample force for a "pseudo-polar" probe is logarithmic in $z_p$ with manifest dependence on the probe size $L$, Eq. (12) nevertheless reveals for this case that $\Delta f \propto 1/z_p$, manifestly independent of $L$ – in other words, $\Delta f$ remains a local probe of magnetism. Therefore, the "magnetic approach curves" $\Delta f(z_p)$ supplied by our magnetic force microscopy measurements encode local information on $r_\mu$, and in principle thereby $\mu$, through Eq. (11).

We utilize the combination of Eqs. (11-12) together with the realistic probe transfer function $\phi_p(q)$ presented in the previous section to synthesize the experimental curves shown in Figs. 3, S6. Fixed-order Gauss-Legendre quadrature is used to integrate Eq. (12) and to evaluate $\Delta f$ at each coordinate $z_p$. The curves $\Delta f(z_p)$ so obtained are then scaled to the dimensional probe-sample distances $z_p$ recorded in the experiment by supposing a reasonable probe radius of curvature $a \approx 20$ nm. The introduction of an *ad hoc* "distance of closest approach" between the probe and sample, so commonplace among quantitative MFM analyses, is completely unnecessary in our formalism. The exponential momentum cutoff characteristic of the hyperboloid probe transfer function (Eq. (5)) naturally supplies such a mathematical feature with no need to impose any additional tuning parameter. We introduce an overall pre-factor $A$ to the computed curves $\Delta f(z_p)$, thus absorbing unknown factors like the thickness of the probe's magnetic layer and its precise magnetization density and cantilever spring constant, which should vary little over the course of our experiments, thus bringing these curves into physical units. Curves of best-fit in comparison to our experimental data are obtained by iterating evaluation of Eq. (1) with respect to $\mu$ using a nonlinear least-squares routine (the Levenberg-Marquardt algorithm), while fixing the probe-dependent pre-factor $A$ to a constant value when fitting across a group of approach curves collected from an individual co-localized map. The value for $A$ for which curves of best fit minimize the average residual with respect to an entire group of experimental curves is the one we select for ultimate fits among that group. Empirically, we find that such choice of $A$ is unique, and varies little between groups of fits. This outcome is expected, since the overall multiplication of force curves by $A$ is an effect linearly independent from the $\mu$-dependence, which rather controls their shape.

## Section 4: Monte Carlo simulations of magnetism in CrSBr

In our model, monolayers of CrSBr are characterized by rectangular lattice structures of $N_R \times N_C$ spins and $N_L$ monolayers stacked and allowed to interact via antiferromagnetic interlayer couplings. The total number of spins is given by $N = N_R \times N_C \times N_L$. The Hamiltonian of this model may be expressed as $H = H_{intra} + H_{inter}$, where the first and second terms represent intralayer and interlayer interactions respectively. The intralayer contribution is given by

$$H_{intra} = \sum_{\langle i,j \rangle} J_{1,x} S_{i,x} S_{j,x} + \sum_{\langle\langle i,j \rangle\rangle} J_{2,x} S_{i,x} S_{j,x} + \sum_{\langle\langle\langle i,j \rangle\rangle\rangle} J_{3,x} S_{i,x} S_{j,x}$$

$$+ \sum_{\langle i,j \rangle} J_{1,y} S_{i,y} S_{j,y} + \sum_{\langle\langle i,j \rangle\rangle} J_{2,y} S_{i,y} S_{j,y} + \sum_{\langle\langle\langle i,j \rangle\rangle\rangle} J_{3,y} S_{i,y} S_{j,y}$$

$$+ \sum_{\langle i,j \rangle} J_{1,z} S_{i,z} S_{j,z} + \sum_{\langle\langle i,j \rangle\rangle} J_{2,z} S_{i,z} S_{j,z} + \sum_{\langle\langle\langle i,j \rangle\rangle\rangle} J_{3,z} S_{i,z} S_{j,z}$$
$$- \sum_i K_x S_{i,x}^2 - \sum_i K_y S_{i,y}^2$$

where $\mathbf{S}_i$ are dimensionless, unit-length vectors representing the magnetic moment at site $i$ and, therefore, the parameters are expressed in units of energy (meV). The sums corresponding to the anisotropic intralayer symmetric exchange parameters $\mathbf{J}_1$, $\mathbf{J}_2$, and $\mathbf{J}_3$ (for which $J < 0$ represents ferromagnetic exchange) are taken over all nearest-, second nearest-, and third nearest-neighbors within the same layer. The parameters $K_x$ and $K_y$ represent single-ion magnetic anisotropy such that $K > 0$ promotes an easy axis. The interlayer contribution to the Hamiltonian is given by

$$H_{inter} = \sum_{\langle i,j \rangle} J_{inter,1} \, \mathbf{S}_i \cdot \mathbf{S}_j + \sum_{\langle\langle i,j \rangle\rangle} J_{inter,2} \, \mathbf{S}_i \cdot \mathbf{S}_j$$

where $J_{inter,1}$ and $J_{inter,2}$ are taken to be the isotropic interlayer exchange parameters between nearest- and second-nearest neighbors between layers. The values of our magnetic parameters are taken primarily from ref. [4], and we choose the interlayer exchange parameters $J_{inter,1} = J_{inter,2} = .012$ meV such that the low-$T$ ground state exhibits antiferromagnet alignment between neighboring layers. The easy axis of magnetization is found to be the $x$-axis, while the $z$-axis is the out-of-plane axis.

In general, calculation of the magnetic susceptibility $\frac{\partial \langle M_z \rangle}{\partial B_z}$, where $\langle M_z \rangle$ is the thermodynamic average of the total magnetization along the $z$-axis, $M_z = \sum_i S_{i,z}$, may be performed by evaluating $\frac{\partial \langle M_z \rangle}{\partial B_z} = \frac{1}{k_B T} [\langle M_z^2 \rangle - \langle M_z \rangle^2]$. Here, we compute the "sheet susceptibility" along several axes which is defined, for example, along the $z$-axis as

$$\chi_{zz} = \frac{N_L}{k_B T \, N} \, [\langle M_z^2 \rangle - \langle M_z \rangle^2]$$

The thermodynamic quantities $\langle M^2 \rangle$ and $\langle M \rangle$ are computed via a single-spin perturbative Metropolis Monte Carlo (MC) algorithm. Lattices of linear dimensions $N_R = N_C = 80$ are annealed to the target temperature $k_B T$ for $N \times 10^5$ MC steps, further equilibration takes place for $2.5 \times N \times 10^5$ MC steps, and observable measurements are recorded over the last $2.5 \times N \times 10^5$ MC steps. The susceptibility values are averaged over an ensemble of $20 - 50$ MC simulations per data point.